\documentclass[useAMS]{mn2e_modmargin}
\usepackage{amsmath}
\usepackage{url}
\usepackage{amsfonts}
\usepackage{amsbsy}
\usepackage{subfigure}
\usepackage{verbatim}
\usepackage{amssymb}
\usepackage{amsbsy}
\usepackage{graphicx}

\renewcommand{\u}{\ensuremath{\mathbf{u}}}
\newcommand{\U}{\ensuremath{\mathbf{U}}}
\renewcommand{\d}{\ensuremath{\partial}}

\newcommand{\ey}{\ensuremath{\mathbf{e}_{y}}}
\newcommand{\ez}{\ensuremath{\mathbf{e}_{z}}}

\title[Dust-gas gravitational instabilities]
 {On dust-gas gravitational instabilities in protoplanetary discs}
\author[H. N. Latter \& R. Rosca]{Henrik N. Latter\thanks{E-mail:
    hl278@cam.ac.uk} \& Roxana Rosca \\
 DAMTP, University of Cambridge, CMS, Wilberforce Road,
Cambridge CB3 0WA, UK}

\begin{document}

\maketitle

\begin{abstract}
In protoplanetary disks the aerodynamical friction between particles and gas
induces a variety of instabilities that facilitate planet
formation. Of these 
we examine the so called `secular gravitational instability' (SGI) in the
two-fluid approximation, deriving analytical expressions for its 
stability criteria and growth rates. 
Concurrently, we present a physical explanation of the instability
that shows how it manifests upon an intermediate range of lengthscales
exhibiting geostrophic balance in the gas component.
In contrast to a single-fluid treatment, the SGI 
is quenched within a critical disk radius, as large as 10 AU and
30 AU for cm and mm sized particles respectively, although establishing robust
estimates is hampered by uncertainties in the parameters
(especially the strength of turbulence) and deficiencies in the
razor-thin disk model we employ. It is unlikely, however, that the SGI is
relevant for well-coupled dust. 
We conclude by applying these results to the question of
planetesimal formation and the provenance of large-scale dust rings.
\end{abstract}


\section{Introduction}
The assembly of planets is a complex and
multi-faceted phenomenon that spans a gulf of some 12 orders of magnitude
in length: from micron-sized dust to $10^3$ km planetary cores. 
It draws on an equally wide range of physical 
processes: collisions, dust-gas aerodynamics, gravitational collapse,
instabilities, and disk structures (e.g.\ vortices, dust traps), 
to name but a few (Papaloizou \& Terquem 2006, Chiang \& Youdin 2010,
Armitage 2010). While it is relatively straightforward to grow cm
sized particles from micron sizes, further growth is potentially 
halted by a number of `barriers' (bouncing, fragmentation, radial-drift;
Johansen et al.~2014). Statistically a small number of `lucky' 
aggregates may hurdle these, but certain collective instabilities
can aid aggregation through this difficult size range. These include
classical gravitational
instability (GI; Safronov 1969, Ward \& Goldreich 1973), streaming instability
(Youdin \& Goodman 2005), and the secular gravitational instability
(SGI; Ward 2000, Youdin 2005). It is to the last instability
 that this paper is devoted.

One of the most attractive features of the single-fluid SGI is that
its onset is unconditional; it should always be present. 
Unlike classical GI, which requires the
Toomre parameter to be less than one, a single fluid analysis presents
no analogous restriction: the SGI works no matter how thin or thick
the particle sub-disk (Youdin 2005). The instability attacks longer scales
preferentially, which ordinarily would be stabilised by the Coriolis force;
but particles can shed (or gain) angular momentum via
areodynamical drag, and hence are not obliged to undergo stabilising
epicycles. As a consequence, rings that are 
radially drifting towards each other continue to do so unimpeded, and the
instability can proceed. On small radial 
scales the SGI is suppressed by dust pressure or gas turbulence, and
in fact, for well-coupled dust,
turbulence decreases growth rates to potentially insignificant levels
(Shariff \& Cuzzi 2011, Youdin 2011). Marginally coupled particles, however,
could still be subject to respectable SGI growth rates at certain
radii. 

The SGI has been thoroughly explored in single fluid models, which are
applicable when the dust to gas density ratio is tiny (e.g.\ Ward
2000, Youdin 2005, Shariff \& Cuzzi 2011, Youdin 2011, Michikoshi et
al.~2012). These models assume that the angular momentum bestowed onto, or
removed from, the gas disk is negligible. On
sufficiently long scales, however, both sides of this momentum transaction must be
included and the gas dynamics explicitly calculated. An
instability criterion then appears:
in a two-fluid model the onset of SGI is no longer
unconditional.
Recently, Takahashi \&
Inutsuka (2014, hereafter TI) made a start on this problem (see also
Shadmehri 2016 and Takahashi \& Inutsuka 2016), 
but there is still much to be
established. Putting aside the
issue of growth rates, an especially important question is: 
at what radii and for what particle sizes
should we expect SGI to exist at all?

The first aim of this paper is to derive clean stability
criteria for the SGI. In the limits of strongly-coupled and
weakly-coupled particles these can be formulated analytically and
involve a variety of parameters, including the gas's Toomre parameter
and the dust-to-gas density ratio. Because they bypass the SGI's full
6th order dispersion relation, these criteria make it relatively
easy to assess its prevalence.

The criteria also motivate a
straightforward physical picture of instability in a
two-fluid system. In order for the instability to work, there must exist
an intermediate range of lengthscales gas upon which (a)
 dust pressure or turbulent mass diffusion is subdominant, and (b) 
the gas is
prevented from executing epicycles, despite its angular momentum
transactions with the dust. Going to lengthscales longer than the
dust pressure (or diffusion) scale takes
care of the first restriction. But the second can only be satisfied if
geostrophic balance holds in the gas fluid, and so we must
simultaneously find shortish scales upon which gas pressure is dominant. The existence
or not of this intermediate range furnishes us with the stability
criterion. 

The formalism is applied to realistic disk models, where we
find that it is unlikely that well-coupled dust is unstable to the SGI
at any radius, unless the background turbulence is especially weak.
Marginally coupled particles, however, can achieve
appreciable growth rates in certain circumstances, emphasising that
the SGI could help aggregation of solids of cm size. We conclude,
however, that SGI is probably unrelated to the dust rings recently
observed by ALMA (Brogan et al.~2015).

The paper will be organised in the following way. First, in Section 2,
we present the two-fluid razor-thin disk model that we employ, alongside a
critical discussion of its shortcomings. The main parameters of the
analysis will also be defined. In Section 3 we revisit the
single-fluid model to fix some ideas and to provide context for the
subsequent analysis, while in Section 4 we briefly treat a simple
two-fluid
system where the gas is regarded as incompressible.
 The main results of the paper are in Section 5, in which we derive
 analytic stability criteria in relevant limits that are then
 applied
to realistic disk models in Section 6. We draw our conclusions in
Section 7, where we discuss the relevance of the
SGI in planet and structure formation in protostellar disks.

\section{Preliminaries}

\subsection{Modelling issues}

The classical GI and secular GI have primarily been explored with
1D models of a vertically averaged or razor-thin disk. Recent notable exceptions are
Mamatsashvili \& Rice (2010) and Lin (2014), who also capture vertical
convection and the magnetorotational instability respectively. A 1D model
certainly eases the analysis and it should be a reasonable
approximation for unstable modes whose radial wavelength is much greater than the disk
thickness; because the classical GI has minimal vertical structure (being
essentially an f-mode in this limit; Ogilvie 1998), 
it is also likely that the SGI  depends on $z$ only weakly. 
For wavelengths closer
to the scale height, a somewhat ad hoc correction may be included
(e.g.\ Shu 1984),
which generally works against instability on these shorter
lengthscales. 

In a two-fluid model, however, the razor-thin assumption is complicated by the
fact that the particle fluid and the gas fluid possess different
thicknesses, 
and the former can be significantly shorter than the latter. 
This is a problem for the SGI
because the fastest lengthscales are not far from the particle scale
height (Youdin 2005), and hence of order or less than the gas scale height. 
As a consequence, the approximation of a razor-thin disk is not
strictly applicable, at least in the description of the gas.
In a single fluid model this issue does
not crop up because it is assumed that the gas fluid is unperturbed by
whatever the dust is doing; but in a two fluid model this is
not the
case.  It may be that the gas perturbations associated with the 
SGI are sufficiently small that the disk's vertical structure plays
little role. But only calculations in vertically stratified shearing boxes
can decide on this point.

A second issue is the correct coupling between the two
fluids. In a real system, with different gas and dust scale heights,
the drag acceleration will be a function of
vertical height $z$. Moreover, 
the entire column of gas will not exert drag on the dust
if the dust subdisk is much thinner. For consistency, the gas
\emph{external} to the dust disk should be excluded from a two-fluid
razor-thin treatment, with the weighting of the drag force
in the momentum equation adjusted appropriately to account for the
smaller surface density of the gas subdisk. Because most of
the mass in both disks is near the midplane, this problem may not 
invalidate the main qualitative results. It should, however, be kept
in mind.
 
A third issue concerns turbulence in the gas, its
effects, and how to mathematically describe it. The
gas is likely undergoing disordered motions at some (perhaps all) radii,
though the underlying physics may differ between different regions
(Turner et al.~2014). The disk may also support turbulence because of
the settling of the dust to the midplane, and ensuing
vertical Kelvin-Helmholtz instability (Cuzzi et al.~1993). The consequences of
turbulence on particles are several. Agitation of the solids
produces enhanced velocity dispersions, over and above that
arising from particle collisions (Goldreich \& Tremaine 1978), and
thus an appreciable particle pressure that will oppose vertical
settling. 
The precise `equation of state'
this pressure obeys, however, is difficult to establish. In addition, the random motions
induced in the dust can potentially smooth away inhomogeneities in the dust
density, and thus lead to diffusion in the continuity
equation directly (Youdin 2011, Shariff \& Cuzzi 2011, Takahashi \&
Inutsuka 2012). 
The efficiency of this diffusion
is pretty much unconstrained and obviously 
requires numerical exploration; for instance, the
vertical Kelvin-Helmholtz instability will mix particles effectively
in the vertical direction but not necessarily in the horizontal.
Finally, turbulence will
transport momentum, though this effect (being a straightforward mild
damping) 
we do not include in this paper (see TI for
its treatment). 

The effects of turbulence on the dust have been mathematically
modelled via mean-field theories, and Langevin and Fokker-Planck
equations (e.g.\ Schr\"apler \& Henning 2004, Carballido et al.~2006, 
Youdin \& Lithwick 2007). Given a Kolmogorov spectrum of isotropic
homogeneous turbulent motions, Youdin \& Lithwick (2007) derive
convenient expressions for the dust velocity dispersion and the
turbulent diffusivity of particle mass, in terms of the stopping time
and a turbulent efficiency parameter (described below). Such a
calculation, of course, must assume that the turbulent flux of
particles behaves as a Fickian diffusion --- which it need not, especially
on scales close to the largest `eddy'. Other complications could
arise from the flux's finite relaxation time and possible
anti-diffusive behaviour 
(especially on small scales, e.g.\ Frisch 1995 Davidson 2000,
Cuzzi et al.~2001). 

Our purpose in this paper is not to improve on any of these issues,
but it is important to flag them at this point. We
persist with the simple models previously employed 
(e.g.\ TI, Shadmehri 2016), mainly because
they can fix ideas and establish clear results, and presumably give
approximately correct predictions.
Future work, however, should involve
vertically stratified models along the lines of Mamatsashvili \& Rice
(2010) and Lin (2014).

\subsection{Parameters}

As will become clear by Section 5, the two-fluid secular GI is
governed by a large number of parameters; it is hence convenient
to define them all in one place. The first key parameter is the 
dust-to-gas mass ratio, denoted by $\delta$ and defined via
\begin{equation}
\delta = \frac{\sigma_a}{\sigma_g},
\end{equation}
where $\sigma_a$ is the background surface density of the dust with
size $a$, and
$\sigma_g$ is the surface density of the gas located \emph{within the dust
subdisk} (see earlier). Thus $\delta$ is a function of $a$.
Because the SGI is
size-selective it is necessary to distinguish between particles of 
different sizes, and thus to use separate surface densities for each
subspecies. 
To ease the analysis in this paper we examine each dust species separately, though
in reality different species will weakly couple via the gas phase. See
Shadmehri (2016) for an attack on a system of interacting gas and two
species of particle.

Generally, $\sigma_a$ is a small fraction of the total
solid surface density. In fact, the surface density of mm and cm
sized particles can be significantly less than 1\% of the total
(e.g.\ Brauer et al.~2008, Windmark et al.~2012a, 2012b),
though this figure varies greatly with age and as different physical processes
are included or neglected
(sticking, bouncing, fragmentation, and mass tranfer, for instance; Garaud
et al.~2013). 
Note that $\sigma_g$ will also be less than the total gas density, as it
only includes gas
situated amongst the dust subdisk. This decrease can be quantified by
a factor $\sim H_g/H$, where $H_g$ and $H$ are the scale thicknesses
of the gas and dust respectively. We may then write
\begin{equation} \label{esty}
\delta \approx 0.01\,\left(\frac{\sigma_a}{\sigma_\text{tot}}\right)\left(\frac{H_g}{H}\right),
\end{equation}  
where $\sigma_\text{tot}$ is the total dust surface density (including
all species), and we have assumed that the ratio of total dust to
total gas density takes the standard value 0.01 (Chiang \& Youdin
2010). Given
the large uncertainties in the second and third factors in
Eq.~\eqref{esty} (small and large respectively), we
simply set $\delta =0.01$ for most calculations.

\begin{figure}
   \centering
   \includegraphics[width=10cm]{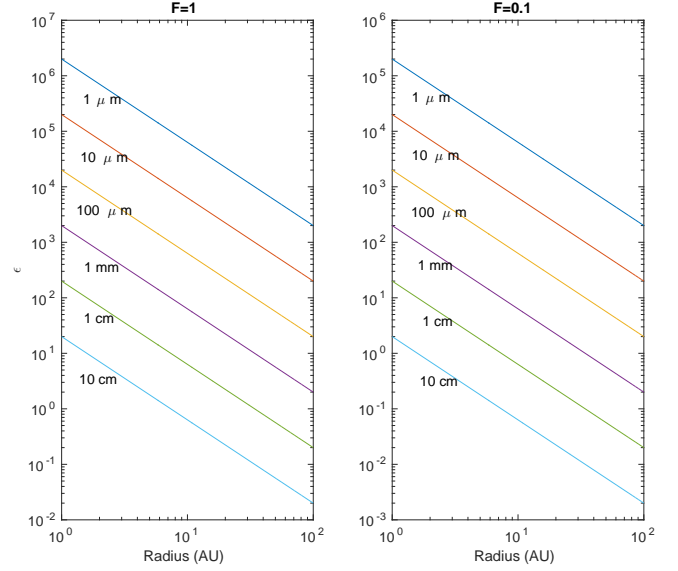}
   \caption{\label{epsilon} Inverse Stokes number $\epsilon$ as a function of radius for
     different
particle sizes and two different minimum mass solar nebulae models,
as calculated from Youdin (2011) but only using the Epstein drag
regime. The $F=1$ case corresponds to a nebula with a gas surface
density of 2000 gcm$^{-2}$ at 1 AU, while $F=0.1$ corresponds to 200
gcm$^{-2}$. See Eq.~\ref{MMSN}.}
\end{figure}

The second key parameter is the inverse Stokes number, which we denote by
$\epsilon$ and define through
\begin{equation}
\epsilon = \frac{1}{\tau_s \Omega},
\end{equation}
where $\tau_s$ is the stopping time of the particles, and $\Omega$ is
the local orbital frequency. Large values of $\epsilon$ correspond,
thus, to strongly coupled particles, and smaller values to weakly
coupled particles. The exact value of $\epsilon$ depends on particle
size, naturally, but also on radial location (e.g.\ Chiang \&
Youdin 2010). 
Fig.~\ref{epsilon}
gives $\epsilon$ profiles for two different minimum mass solar nebulae
(MMSN) models. See Section 6 for more details.

The third parameter is the ratio
of dust and gas velocity dispersions, defined through
\begin{equation}
\eta = \frac{c^2}{c_g^2},
\end{equation}
where $c$ and $c_g$ denote the velocity dispersion of the dust and gas
respectively. To a first approximation, the dust random velocities are
controlled by `kicks' delivered by the gas turbulence, rather than
inter-particle collisions (Youdin \& Lithwick 2007). Thus it is
possible to relate $\eta$ to properties of the turbulence, and we do
so below. Collisional agitation becomes important for 
larger particles $\gtrsim$ cm, and thus the $\eta$ we use in this
paper may be an underestimate for these sizes.

The fourth parameter is the Toomre $Q$ of the dust, which describes the
onset of classical GI. It is given by
\begin{equation}
Q = \frac{c\Omega}{\pi G\sigma},
\end{equation}
where $G$ is the gravitation constant and for notational convenience
we have dropped the subscript `$a$'
from the surface density. An analogous expression can be
defined for the gas, which we denote by $Q_g$. The two Toomre
parameters can be related via the following identity: $Q_g=
(\delta/\sqrt{\eta})Q$. 

Finally, the turbulent diffusion of solid particles can be quantified
via the mass diffusivity $D$, and the diffusion of gas by the analogous
$D_g$. As in Youdin (2011), we replace these by the dimensionless parameters $\alpha$
and $\alpha_g$, via $D=\alpha c_g^2/\Omega$ and $D_g=\alpha_g
c^2_g/\Omega$, with these related via
\begin{equation}
\alpha =
\frac{\epsilon^2+\epsilon+4}{(\epsilon+\epsilon^{-1})^2}\alpha_g,
\end{equation}
(Youdin \& Lithwick 2007, Youdin 2011). The dust's velocity dispersion
(if excited by turbulence) can also be expressed in terms of $\alpha_g$
and we find
\begin{equation}
\eta = \frac{\epsilon^3+2\epsilon + 5/4}{\epsilon(\epsilon+\epsilon^{-1})^2}\alpha_g.
\end{equation}
Thus $\alpha_g$ governs both $D$ and $\eta$. Unfortunately, an
estimate of its magnitude is one of the great uncertainties in the
theory, though it is likely to be smaller (and possible much smaller)
than the analogous dimensionless coefficient associated with angular
momentum transport in the gas ($10^{-3}-10^{-2}$ in a protoplanetary
disk). For instance, if transport is controlled by
magnetocentrifugal winds or strong zonal magnetic fields 
then the bulk of the disk could even be laminar 
(e.g.\ Lesur et al.~2014, Bai 2014). Having said that, vertical
settling should always lead to disordered motions and some degree
of radial diffusion.
Putting these considerations aside, it is clear that strongly coupled
particles ($\epsilon\gg 1$) have $\alpha\sim \alpha_g$ and $\eta\sim \alpha_g$;
whereas weakly coupled particles ($\epsilon \ll 1$) are diffused less effectively
$\alpha\sim\epsilon^2\alpha_g$ and are far `colder', $\eta \sim
\epsilon\alpha_g$. 

\section{A single fluid model}

Though the single fluid analysis of the SGI is well-trodden territory
we include it here for completeness and because it helps fix
useful ideas that appear later. One especially important theme that
crops up is geostrophic balance and the purely azimuthal `zonal'
flows that ensue.

\subsection{Governing equations}

As we are interested in relatively short radial scales it is
convenient to employ the shearing sheet model (Goldreich \&
Lynden-Bell 1965), whereby a small portion
of disk, centred upon a fixed radius $R_0$, 
is represented by a Cartesian box.
In a corotating frame centred on the box, 
$x$ and $y$ denote the local radial and
azimuthal coordinates, while $\Omega$ is the orbital frequency
at $R_0$. The disk is assumed to be razor thin, so that the dust
volumetric density is $\sigma(x,y)\delta(z)$, where $\sigma$ is the
dust surface density and $\delta(z)$ the Dirac delta function (not the
dust-to-gas mass ratio). 

The equations governing the evolution of the dust fluid are given by
the continuity, momentum, and Poisson equations:
\begin{align}
&\d_t \sigma + \u\cdot \nabla\sigma = -\sigma\nabla\cdot\u +D\d_x^2\sigma, \label{dust1}\\
&\d_t\u +\u\cdot\nabla\u = -\nabla\Phi_t -\nabla\Phi_\text{sg}-\nabla
P \notag \\& \hskip3cm - \epsilon\Omega (\u- \mathbf{U}) -2\Omega\ez\times\u, \\
& \nabla^2 \Phi_\text{sg} = 4\pi G\sigma\,\delta(z),\label{dust2}
\end{align}
where $\sigma$, $\u$, and $P$ denote the dust surface density,
velocity, and pressure. Again
we have dropped the subscript `$a$'
from the surface density; it is understood from here onwards
 that $\sigma$ refers to the surface density of particles of size $a$.
 The potentials associated with the dust
self-gravity and tide are $\Phi_\text{sg}$ and
$\Phi_t=-(3/2)\Omega^2x^2$ respectively.
The gas velocity is given by
$\mathbf{U}$, and it interacts with the dust via a drag term whose
strength is quantified by the inverse Stokes number $\epsilon$.

In this section it is
assumed that there is no appreciable backreaction of the dust on the
gas motion. Moreover, we neglect the effect of any radial pressure
gradient on the gas's orbital rotation.
It is thus Keplerian and steady:
$\mathbf{U}=-(3/2)\Omega \ey$. In realistic disks there is likely to
be a fluctuating component of the gas motion due to turbulence, which
acts as a forcing term in the dust momentum equation, giving rise
to velocity fluctuations in the dust. We will be interested in
the large-scale mean dust velocity,
 rather than these fluctuations; the latter's effects
may be captured by a turbulent pressure tensor in the momentum
equation and a turbulent radial flux in the continuity equation.
The $\nabla P$ and $D\d_x^2\sigma$ terms are the manifestations of these
two effects. 

Finally, given that we have assumed a dust pressure, we must stipulate
the dust's equation of state, relating $\sigma$, particle
velocity dispersion $c$, and $P$. This is not straightforward, especially if the
dust velocity dispersion is dominated by the turbulent gas
fluctuations. As our intention is to provide a broad physical explanation
of underlying physics, rather than detailed modelling, we assume for
simplicity that the dust fluid is isothermal, and so $P=c^2 \sigma$,
where $c$ is constant.

\subsection{Dispersion relation}

The governing one-fluid equations admit an equilibrium characterised
by a constant density $\sigma=\sigma_0$ and the perfect entrainment of the dust in the
gas, both undergoing Keplerian motion $\u=\u_0=
-(3/2)\Omega\,x\ey$. To this steady state we add small axisymmetric
perturbation,
$\sigma'$, $\u'$, proportional to $\text{exp}(st + \text{i}kx)$, where
$s$ is a growth rate and $k$ is a radial wavenumber. Their
linearised equations are
\begin{align}
s\sigma' &= -\sigma_0 \text{i}k u_x', \label{lin1}\\
su_x' &= 2\Omega u_y' - \text{i}k\Phi_\text{sg}' -
c^2\text{i}k\frac{\sigma'}{\sigma_0}-\epsilon\Omega u_x', \label{lin2}\\
su_y' &= -\tfrac{1}{2}\Omega u_x' -\epsilon\Omega u_y', \label{lin3}
\end{align}
where the perturbed gravitational potential is given by
$\Phi_\text{sg}' = - (2\pi G/|k|)\sigma'$
(e.g., Binney \& Tremaine 1987) and where we have set $D=0$ for the
time being. 

Eliminating the primed variables produces a relatively neat 
third order dispersion relation:
\begin{align} \label{1Fdisp}
s^3 + 2\epsilon\Omega s^2 + (\overline{\omega}^2 +\epsilon^2\Omega^2)s
+\epsilon\Omega(\overline{\omega}^2-\Omega^2)=0.
\end{align}
Here 
\begin{equation} \label{ombar}
\overline{\omega}^2 = \Omega^2 -2\pi G\sigma_0|k|+k^2c^2,
\end{equation}
 is
the standard expression for the squared frequency of density waves in
a 1D inviscid disk. This agrees with multiple examples in the
literature, notably in Ward (2000) and Youdin (2005), and also in Youdin
(2011), and Shareef \& Cuzzi (2011), when turbulent diffusion is
omitted. 

\subsection{Without gas drag}

It is worthwhile examining the classical case with no drag,
i.e. when $\epsilon=0$. The dispersion simplifies and one obtains $s=
\pm \text{i}\overline{\omega}$ and $s=0$. The first pair of solutions
corresponds to 1D density waves, which can grow on a band of
intermediate wavenumbers $k$ girdling 
$$k_c= \pi G\sigma_0/c^2.$$ 
The
instability criterion requires the Toomre parameter $Q= \Omega
c/(\pi G\sigma_0)$ to be less than 1 (e.g.\ Safronov 1969, Goldreich \& Ward
1973). 
Radial collapse on long wavelengths is impeded by epicyclic motion induced by
the inertial forces, whereas short wavelengths
modes are stabilised by
pressure. (Note that nonlinear non-axisymmetric instability
occurs for larger $Q\approx 2$.) The dust, however, must be rather
thin in order to achieve
$Q<1$ (e.g.\ Cuzzi et al.~1993, Chiang \& Youdin 2010) 

The $s=0$ `quasi-geostrophic' mode neither grows nor oscillates, but instead
corresponds to a steady `zonal flow'. From Eq.\eqref{lin2}, the
fundamental balance is between the Coriolis force, on one hand, and
self-gravity and pressure, on the other. The mode corresponds to a
radially varying sequence of super and sub-Keplerian orbital motions
(or `jets'):
\begin{equation} \label{geostroph}
u_y' = \frac{\text{i}k}{2\Omega}\left(1-2\frac{k_c}{|k|}\right)P',
\end{equation}
where $P'$ is the associated pressure perturbation. The pressure
gradient negates any type of epicyclic motion. This type of flow plays an
important part in the secular GI, as will
be made clear in the following sections, and indeed in the streaming
instability (Jacquet et al.~2011).

\begin{figure}
   \centering
   \includegraphics[width=9cm]{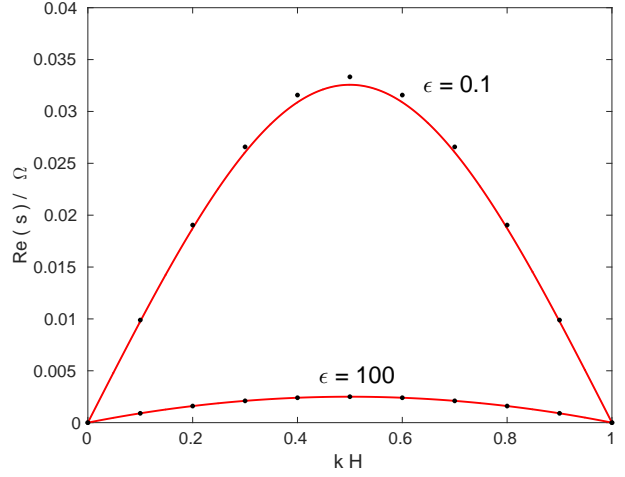}
   \caption{\label{fig1} Growth rates of the secular GI in a single
     fluid for $Q=2$ and two different values of $\epsilon$, 0.1
     (weakly coupled)  and
     100 (strongly coupled).
     The solid curves represent the
   full solution to the dispersion relation \eqref{1Fdisp}, whereas
   the points correspond to the asymptotic estimates
    \eqref{epl} and \eqref{epg}.}
\end{figure}

\subsection{With gas drag}

Being a cubic, the dispersion relation \eqref{1Fdisp} does not
yield simple formulae for the various growth rates. It is
straightforward, however, to obtain a general stability criterion and
 asymptotic expressions in the
limit of strong and weak drag. 

The last term in the cubic reveals that linear instability is assured
when $\overline{\omega}^2-\Omega^2 <0$, which is satisifed for all
modes with sufficiently small wavenumbers $k< 2k_c$. Hence instability
is unconditional, though in practice (putting aside the disk's global
structure and cylindrical geometric effects) an unstable dust layer must
have radial extent greater than $\approx 2\pi/k_c$ or else there will
insufficient space for the modes to manifest themselves. 

In Figure \ref{fig1} we plot numerical growth rates of the secular GI for
parameters characteristic of the two limits. Superimposed are the
leading order estimates taken from Eqs \eqref{epl} and \eqref{epg} below.

\subsubsection{Weak coupling limit}

When $\epsilon \ll 1$ we are in the weak drag regime corresponding to larger
particles, at larger disk radii, in less massive disks.
 Eq.\eqref{1Fdisp} then yields two density waves:
\begin{equation}
s= \pm\text{i}\,\overline{\omega}
-\frac{1}{2}\Omega\epsilon\left(1+\frac{\Omega^2}{\overline{\omega}^2}\right)
+\mathcal{O}(\epsilon^2\Omega),
\end{equation}
which are unstable, as expected, when $Q<1$ and thus correspond to the
classical GI. Otherwise the two waves are mildly
damped by drag. 

The third
secular mode exhibits a growth rate of
\begin{equation} \label{epl}
s =
\frac{\Omega^2-\overline{\omega}^2}{\overline{\omega}^2}\epsilon\Omega
+ \mathcal{O}(\Omega\epsilon^2).
\end{equation}
Instability is assured when $\overline{\omega}^2<\Omega^2$, in
accordance with the stability criterion derived above. 
 The fastest
growing mode possesses $k=k_c$ and the maximum growth rate is
\begin{equation} \label{1fwcs}
s_\text{max} \approx \frac{\epsilon\,\Omega}{Q^2-1}.
\end{equation}
The mode always grows, but for large $Q$ the growth rate can be small.

 In the limit of small $\epsilon$, the
unstable mode is, in fact, a modified zonal flow. In Eq.\eqref{lin2}, the
dominant balance is \emph{quasi-geostrophic}, i.e. 
between the pressure gradient, self-gravity, and
the Coriolis force (because $s\sim\epsilon\Omega$ and $u_x\sim
\epsilon u_y$). Using Eq.~\eqref{geostroph}, with Eqs \eqref{lin1} and
\eqref{lin3}, gives then precisely the leading order term for $s$.  

\subsubsection{Strong coupling regime}

In the strong drag limit, $\epsilon \gg 1$,
the SGI grows at a rate of
\begin{equation}\label{epg}
s= \frac{\Omega^2-\overline{\omega}^2}{\Omega}\,\epsilon^{-1}+\mathcal{O}(\Omega\epsilon^{-2}).
\end{equation}
Instability occurs for the exact same range of $k$ as in the opposed
weak drag limit, but the maximum growth rate is slightly altered
\begin{equation}
s_\text{max} \approx \frac{\Omega}{Q^2\epsilon}.
\end{equation}
and the mode works through a different arrangement of forces. In
Eq.~\eqref{lin2} instead of radial geostrophic balance, it is the last
three terms that are dominant: radial contraction by self-gravity is met
by the drag force and pressure. Essentially, the dust has achieved \emph{terminal
velocity}. Solving for $u_x'$ and using \eqref{lin1} yields the leading
term for $s$. 

In addition, there exist two density waves with
$s=\pm\text{i}\overline{\omega}+\mathcal{O}(\epsilon^{-1}\Omega)$. 
As in the weak coupling limit, they grow exponentially when $Q<1$ but are
otherwise weakly damped by drag at a rate $\sim \Omega/\epsilon$.

\subsubsection{Role of turbulent diffusion}

For completeness we next consider the impact of turbulent mass
diffusivity, focussing especially on the critical wavenumber at
which instability is quenched. Letting $D\neq 0$, and 
taking the limit of weak drag, $\epsilon\ll
1$, the leading order expression for the growth rate \eqref{epl} picks up a
term $-k^2 D\Omega^2/\overline{\omega}^2$.
The critical $k$ is then easy to compute:
\begin{equation}
k_\text{crit} = \frac{2\pi G\sigma_0}{c^2+D\Omega/\epsilon}.
\end{equation}
The importance of turbulent diffusion versus pressure 
is quantified by $D\Omega/(\epsilon c^2)$. 
For weakly coupled particles, this parameter
is $\sim \alpha(\eta \epsilon)^{-1}\sim 1$, using the estimates on
turbulent diffusion and velocity dispersion from Section 2. This is
saying that
turbulence is roughly as important as dust pressure, at least
in setting the short scale cut-off for SGI. Nonetheless, in the remainder of the
paper we often omit diffusion when dealing with weakly coupled
particles, primarily in order to derive clean expressions. The neglect
of diffusion will not change these results qualitatively, but will
introduce order 1 corrections. Instability criteria should then
be regarded as necessary conditions, not sufficient, and maximum
growth rates should be understood as upper bounds.

In the opposite limit, $\epsilon\gg 1$, the situation is quite different. 
The
growth rate \eqref{epg} is modified by the term $-k^2D$, and
the importance
of turbulent diffusion on small scales is quantified instead by $\epsilon D\Omega/c^2$
which is $\sim \epsilon \gg 1$. As a consequence,  the critical
cut-off for small particles
is completely controlled by turbulence, as shown earlier by
Youdin (2011) using a heuristic (but essentially equivalent)
argument.

On account of the subdominance of particle pressure in the face of
turbulent diffusion when $\epsilon\gg 1$, we may dispense with it
entirely. In this case,
instability is assured for sufficiently long
wavelengths as earlier, with the asymptotic growth given by
\begin{equation} \label{pless1f}
s = \frac{2\pi G\sigma_0}{\epsilon\Omega}k - Dk^2,
\end{equation}
and a maximum growth rate of
\begin{equation}\label{pless1fmax}
s_\text{max} = \frac{\pi^2 G^2 \sigma_0^2}{\epsilon^2\Omega^2 D}\approx
\frac{\Omega}{\epsilon^2\alpha_g Q_g^2},
\end{equation}
where we have introduced the gas Toomre parameter in the last
equality, and set $\alpha\approx \alpha_g$ (cf.\ Section 2.2).
The wavelength of maximum growth is
$\sim \epsilon\alpha_g Q_g H_g$, where $H_g$ is the scale height of
the gas disk. Given that in this regime $\epsilon$ is large but
$\alpha_g$ is small, it can be hard to establish this characteristic
length a priori.

Before moving on, it should be highlighted that a turbulent
mass-diffusion alters the classical GI in unexpected ways. 
When $\epsilon=0$ and $D\neq 0$, 
arbitrarily long wavelengths are rendered unstable, though they grow
at negligible rates. On the other hand, mass diffusion cannot
completely stabilise short scales - a pressureless turbulent fluid
will be unstable for all $k$ (in contrast to the SGI). Introducing turbulent
momentum diffusion, however, does kill instability for
sufficiently large $k$. We omit details of these calculations.

\subsection{Physical picture}

As has been commented upon in the literature, the secular GI exhibits two striking
features: growth for all finite $Q$, and growth upon arbitrarily small $k$. The
latter is especially unexpected given that traditional GI prefers
intermediate wavelengths, the longer scales stabilised by the dominant
Coriolis force. How does secular GI overcome the epicyclic response
at large scales? 

Consider two dust rings located at different radii undergoing
circular orbital motion. Each ring
contains a quantity of angular momentum naturally associated with its home
radius. Suppose the rings are displaced radially towards one another
other. Though
their mutual self-gravitational attraction will attempt to amplify
the displacement, the two rings possess an angular momentum
incommensurate with their new radial location and hence undergo
epicyclic oscillations that thwart any type of gravitational collapse.
This is the classical picture of stabilisation at long wavelengths.

Suppose however that there exists a drag force on both dust rings due to
interactions with the background gas. Now when the two rings are
radially displaced they will exchange angular momentum with the gas, via the
last term in Eq.~\eqref{lin3}. The rings either gain or lose angular
momentum until they achieve the amount commensurate with their
new radial location. As a result, they do \emph{not} undergo epicyclic
motion, and self-gravity continues to amplify 
their radial drift towards one another. 
As the two rings collapse they continuously shed or gain angular momentum as needed.
In this way, the epicyclic restoring forces are negated by the drag.

Note that this scenario only works if the gas remains an infinite
reservoir of angular momentum, which can be removed or added
to with no ramifications. This may be a reasonable approximation in
cases where the dust density is far less than the gas density, but
on some sufficiently long scale even this must break down. The
question then arises: on what range of scales does the
secular GI operate upon, and under what circumstances may we take the
single fluid approximation? For the smooth running of the SGI,
the gas fluid must resist undergoing epicyclic oscillation when
perturbed by the dust drag. When and how can this be arranged?
These questions will be explored in the
following two sections.

\section{Two-fluid model: incompressible gas}

As an intermediate step between the single fluid and fully
compressible 
two-fluid models, we briefly analyse the case of a compressible dust disk
embedded in an incompressible gas. This situation mimics the case when
the dust scale height is far less than the gas scale height, and 
 the unstable motions very
subsonic. On the vertical scale of the dust disk, the gas
density is effectively constant and the problem is
`vertically local' as far as the gas is concerned. Consequently, the
gas density does not contribute to the perturbed Poisson equation.  

\subsection{Governing equations}

The equations governing the evolution of the incompressible gas are
\begin{align}
\d_t\U + \U\cdot\nabla\U &= -\nabla P_g/\sigma_g -\nabla\Phi_t
-\nabla\Phi_\text{sg}\notag \\
& \hskip1cm -2\Omega\ez\times\U + \epsilon\Omega\frac{\sigma}{\sigma_g}(\u-\U),
\\
\nabla\cdot\U&=0,
\end{align}
where $\sigma_g$ and $P_g$ are the gas surface density and vertically
integrated pressure, \emph{within the dust layer}. Because the gas is
incompressible $\sigma_g$ is a constant. As earlier, $\U$ is
the gas velocity, and $\u$ is the dust velocity. The equations
governing the dust fluid are those that appear in Section 2, Eqs
\eqref{dust1}-\eqref{dust2}. To ease the analysis $D=0$.

\subsection{Dispersion relation}

Once again, we assume the standard equilibrium state
$P_g=P_{g0}$, $\sigma=\sigma_0$, $\u=\U=-(3/2)\Omega\,x\ey$,
where the gas pressure and dust surface density is constant.
Next, axisymmetric perturbations are assumed, denoted by 
$\U'$, $P_g'$, $\u'$, $\sigma'$, and these are taken to be $\propto
\text{exp}(st+\text{i}kx)$. 

Because of the incompressibility
condition we immediately obtain $U_x'=0$, and $U_y'$ is computed from
the $y$-force balance, yielding
\begin{equation} \label{Uy}
 U_y' = \frac{\epsilon\,\delta\,\Omega}{s+\epsilon\,\delta\Omega}u_y',
\end{equation}
where we have introduced now
$\delta=\sigma_0/\sigma_g$ which quantifies the dust-to-gas density
ratio for a given particle size.
This equation states that the gas is azimuthally accelerated
by dust drag. Simultaneously, a form of radial geostrophic balance
holds for the gas, with the Coriolis force balanced by the radial
pressure pressure, self-gravity, and drag. Importantly, the gas
pertubation cannot undergo epicyclic motion, which might impede the
growth of the secular GI. The gas pressure gradient holds the fluid
radially `in place' and a sequence of azimuthal jets ensues, each accelerated by
the dust drag. 
(Note that the absence of epicycles is a generic feature of
incompressible flow confined to the orbital plane.)

The perturbation equations
for the dust are the same as Eqs \eqref{lin1}-\eqref{lin3} except for
the inclusion of the term $\epsilon\Omega U_y'$ on the right side of the dust's
$y$-force balance. Eliminating the dependent variables obtains the following
quartic dispersion relation:
\begin{align}
&s^4 + (2+\delta)\epsilon\Omega\,s^3 + [\overline{\omega}^2+
(1+3\delta)\epsilon^2\Omega^2]s^2 \notag \\
& \hskip2cm
+\epsilon\Omega\left[\delta(\overline{\omega}^2+2\epsilon^2\Omega^2)
+\overline{\omega}^2-\Omega^2\right]s \notag\\
&\hskip4cm+2\delta(\overline{\omega}^2-\Omega^2)\epsilon^2\Omega^2=0,\label{2Fincdisp}
\end{align}
which we now briefly analyse.

\subsection{Instability criterion and asymptotic growth rates}

Though Eq.~\eqref{2Fincdisp} may appear rather formidable, an instability
criterion appears immediately. Putting aside the clasical GI for now, 
the SGI mode is marginal when
the last term is zero, yielding exactly the same instability criterion 
as in the single fluid model: $\overline{\omega}^2 <\Omega^2$, and
hence instability occurs on all $k<2k_c$. Though
we allow for gas perturbations, gas incompressibility restricts these
to a form of zonal flow which absorbs or bestows angular momentum as
necessary to faciliate instability in the dust. 

In the weak coupling limit, $\epsilon\ll 1$, the leading order term in
the SGI growth rate is obtained by setting $s= s_1\epsilon + \dots$ and
substituting this into \eqref{2Fincdisp}. One obtains the quadratic:
\begin{equation}
s_1^2 +\Omega(\delta-\xi)s_1 -
2\delta\frac{(\Omega^2-\overline{\omega}^2)}{\overline{\omega}^2}\Omega^2 = 0.
\end{equation} 
The resulting solution for $s_1$ agrees with the single fluid 
expression \eqref{epl} to leading order in small dust-to-gas fraction 
$\delta<1$. The reason for the dependence on $\delta$ is because the 
ability to transfer angular momentum between gas and dust is
influenced by the relative azimuthal speeds of the two fluids, which 
depends on $\delta$ via \eqref{Uy}.

In the strong coupling limit, $\epsilon\gg 1$,
assuming that $s\sim \Omega/\epsilon$, a similar analysis reveals that
the SGI growth does not depend on $\delta$ at all. In fact,
$s = (\Omega^2-\overline{\omega}^2)/(\epsilon\Omega)$, precisely the
same expression as in the single fluid case \eqref{epg}.

\section{Two-fluid model: compressible gas}

Having treated simpler models of the dust-gas system, we 
turn to a fully compressible two-fluid approach, informed by what we have
learned so far. The sound speed of the
gas $c_g$ and its scale height $H_g=c_g/\Omega$ are assumed finite,
with the dust subdisk embedded in the gas, so that
$c<c_g$ and $H<H_g$. As in the previous section, we average over the
vertical thickness of the dust disk, and thus neglect complications
arising on the smaller scales $<H$, such as shear instabilities and
turbulence. These are included in an ad hoc way,
using a turbulent mass diffusivity and enhanced dust pressure. 
Perhaps more importantly, the gas outside the dust disk is completely
neglected as far as the onset of instability is concerned. The
external gas is `inert' --- both gravitationally and dynamically
decoupled. The resulting model is workable but suffers the
shortcomings discussed previously in Section 2.

\subsection{Governing equations}

We adopt the equations listed in
TI in order to describe our system:
\begin{align}
\d_t\sigma + \u\cdot\nabla\sigma &= -\sigma\nabla\cdot\u +
D\nabla^2\sigma, \\
\d_t\u + \u\cdot\nabla\u &= -\frac{1}{\sigma}\nabla P -\nabla\Phi_t 
-\nabla\Phi_\text{sg} \notag \\
&\hskip1cm -2\Omega\ez\times\u +\epsilon\Omega(\U-\u), \\
\d_t\sigma_g + \U\cdot\nabla\sigma_g &= -\sigma_g\nabla\cdot\U, \\
\d_t\U + \U\cdot\nabla\U &= -\frac{1}{\sigma_g}\nabla P_g -\nabla\Phi_t 
-\nabla\Phi_\text{sg} \notag \\
& \hskip1cm -2\Omega\ez\times\U
+\epsilon\Omega\frac{\sigma}{\sigma_g}(\u-\U), \\
\nabla^2\Phi_\text{sg} &= 4\pi G(\sigma+\sigma_g)\delta(z).
\end{align}
Both dust and gas are assumed isothermal, so that $P=\sigma c^2$ and
$P_g=\sigma_g c_g^2$. 
As mentioned, 
by assuming that the dust and gas enclosed in $|z|<H$ is razor thin,
we omit the gravitational influence of gas external to the dust disk.
As a consequence, $\sigma_g$ should be understood as the surface
density of the gas located within the vertical extent of the dust subdisk.

\subsection{Dispersion relation}

Once again we assume a simple background equilibrium of homogeneous
density and Keplerian shear: $\sigma=\sigma_0$,
$\sigma_g=\sigma_{g0}$, $\u=\U=-(3/2)\Omega x \ey$. To this we 
add disturbances, denoted by primes, that are $\propto
\text{exp}(st+\text{i}kx)$. The resulting linearised equations are:
\begin{align}
s \sigma' &=-\sigma_0\text{i}k u_x' - Dk^2 \sigma', \label{2flin1}\\
su_x' &= -\text{i}k c^2(\sigma'/\sigma_0) + 2\Omega u_y' -\text{i}k\Phi'
+\epsilon\Omega(U_x'-u_x'),  \label{2flin1b}\\
s u_y' &= -\tfrac{1}{2}\Omega u_x' + \epsilon\Omega (U_y'-u_y'), \\
s \sigma_g' &= -\sigma_{g0}\text{i}k U_x', \\
sU_x' &= -\text{i}k c_g^2(\sigma_g'/\sigma_{g0}) +2\Omega U_y'-
\text{i}k\Phi'+\epsilon\delta\Omega(u_x'-U_x'), \label{2flin1c} \\
sU_y' &= -\tfrac{1}{2}\Omega U_x' + \epsilon\delta\Omega(u_y'-U_y'), \label{2flin2}
\end{align}
where the equilibrium dust-to-gas density ratio is
$\delta=\sigma_0/\sigma_{g0}$ for a given size $a$. 
Finally, the perturbed gravitational potential is obtained from
$$ \Phi'= -(2\pi G/|k|)(\sigma'+\sigma_g'). $$

The system of equations \eqref{2flin1}-\eqref{2flin2} yields a
rather involved 6th order dispersion relation:
\begin{equation} \label{2fbig}
s^6+a_5 s^5+a_4 s^4 + a_3 s^3 + a_2 s^2 + a_1 s  + a_0 =0,
\end{equation}
with
\begin{align*}
a_5&=2(1+\delta)\epsilon\Omega + D k^2, \\
a_4&=
(1+\delta)^2\epsilon^2\Omega^2+\overline{\omega}^2+\overline{\omega}^2_g
+2(1+\delta)\epsilon\Omega D k^2, \\
a_3&=
\epsilon\Omega\left[(1+\delta)2\mu^2 - \delta
  c_g^2k^2+(1+2\delta)c^2k^2\right] \\
& \hskip2cm +\left[(1+\delta)^2\epsilon\Omega^2+\overline{\omega}^2_g +
\Omega^2\right]Dk^2, \\
a_2 &= c^2k^2\overline{\omega}_g^2+\Omega\mu^2-2c_g^2k^2\pi G\sigma_0|k|
\\
& \hskip1cm +(1+\delta)\epsilon^2\Omega^2\left[(1+\delta)\mu^2+\delta
  c^2k^2-\delta c_g^2 k^2\right] \\
& \hskip2cm +\left[2(1+\delta)\overline{\omega}_g^2-c_g^2k^2 \right]
\epsilon\Omega D k^2, \\
 a_1 &=  \epsilon\Omega k^2\left[\delta
   c_g^2(\mu^2-c_g^2k^2)+c^2(\mu^2+\delta c_g^2 k^2) \right] \\
& \hskip1cm
+\left\{\overline{\omega}_g^2\left[1+\epsilon^2(1+\delta)^2\right] 
   -\epsilon^2\delta(1+\delta)c_g^2k^2 \right\}Dk^2\Omega^2,\\
a_0 &= D c_g^2 k^4\delta\epsilon\Omega^3.
\end{align*}
In order to ease the presentation of the coefficients we have introduced the following
frequencies
\begin{align*}
 \overline{\omega}_g^2&=\Omega^2-2\pi G\sigma_{g0}|k|+c_g^2k^2, \\
\mu^2&=\Omega^2-2\pi G(\sigma_{g0}+\sigma_0)|k|+c_g^2k^2.
\end{align*}
Recall that $\overline{\omega}^2$ is the squared frequency of density
waves in the dust fluid given by Eq.~\eqref{ombar}.

\subsection{No turbulent mass diffusion}

In this subsection we analyse the case when mass diffusion in the
continuity equation is negligible, $D\approx 0$, but we retain
the velocity dispersion of the dust particles. This situation may
adequately describe a disk of weakly or marginally coupled dust and
gas, $\epsilon\lesssim 1$. For this case the dust disk is expected to
be thinner and
`colder' than the gas disk, and so $\eta = c^2/c_g^2 \ll 1$. However,
because of the omission of mass diffusion, the instability conditions
derived here should
be regarded as necessary, not sufficient, and maximum
growth rates understood as upper bounds.

\subsubsection{Instability criterion}

When $D=0$ the dispersion relation \eqref{2fbig} 
reduces to a quintic. Marginality corresponds then to $a_1=0$, which
provides us with a condition for the onset of instability.
 In fact
$a_1=0$ is a quadratic equation for $k$, 
\begin{equation}\label{quad1}
 c^2c_g^2 k^2-2\pi G\sigma_{g0} (c^2+\delta c_g^2)|k| +(c^2+c_g^2\delta)/(1+\delta) = 0, 
\end{equation}
indicating that if
instability occurs it takes place on a range of wavenumbers
$k_1 < k < k_2 $, where $k_1$ and $k_2$ are solutions to the
quadratic. In order for such a range to exist, the critical $k$'s must
be real. This is the case when the discriminant of \eqref{quad1} is
positive, and the instability criterion proceeds easily:
\begin{equation} \label{stab1}
Q_g \equiv \frac{\Omega c_g}{\pi G \sigma_{g0}}<
\sqrt{\frac{(1+\delta)(\delta+\eta)}{\eta}}
 \lesssim  \sqrt{1+\frac{\delta}{\eta}},
\end{equation}
where the last scaling arises if $\delta<1$.
Alternatively, the criterion can be reworked in terms of the dust's
Toomre parameter, noting that $Q= \sqrt{\eta}\delta^{-1}Q_g$. We then
obtain instability when
\begin{equation}
Q \lesssim \frac{\sqrt{\delta+\eta}}{\delta} \approx \delta^{-1/2},
\end{equation}
where in the last approximation we assume that $\eta\ll\delta$ 
(always the case, unless turbulence is
absent--- see Section 2.2).

However expressed, this simple criterion encapsulates clearly the main
physical effects. For instance, if we take the limit of negligible
dust, $\delta\to 0$, the instability criterion
become simply $Q<\infty$, and the system is always
unstable. In this limit the dust's drag on the gas is unimportant and
the the two-fluid system reproduces the same stability behaviour as the
single fluid model (Section 3): the Toomre parameter does not feature.

For general $\delta$, however, stability in the two-fluid disk does in
fact depend on the gas's (or dust's) Toomre parameter: if it is too
large then instability is switched off.
If $\delta/\eta \gg 1$, 
the instability criterion may be written as
\begin{equation} \label{stabbers1}
Q_g \lesssim \frac{c_g}{c}\left(\frac{\sigma_0}{\sigma_{g0}}\right)^{1/2}
\approx \left(\epsilon\alpha_g\right)^{-1/2}\left(\frac{\sigma_0}{\sigma_{g0}}\right)^{1/2},
\end{equation} 
which may greatly exceed the classical value of 1, allowing
instability to occur in a finite range of conditions. In this case
we recover the secular GI.

\subsubsection{Asymptotic growth rates}

Explicit expressions for the growth rate are possible in the limit of 
$\epsilon \ll 1$. We set aside the classical GI and
 isolate the SGI mode
by setting $s= s_1\epsilon+\dots$. Collecting
terms of order $\epsilon$, we obtain
\begin{equation}\label{asymp2f1}
s_1 =
\left(\frac{\sigma_0}{\sigma_{g0}}\right)\left(\frac{2\pi
    G\sigma_{g0}|k|
-\Omega^2}{\overline{\omega}_g^2}\right)\left(\frac{k^2c_g^2}{\Omega}\right)
- \frac{k^2c^2}{\Omega},
\end{equation}
to leading order in small $\delta$ and $\eta$. The last damping term
arises from the particle pressure and kicks in at short scales, of order
$H=c/\Omega$. Instability is thus
restricted to scales longer than $H$. Conversely, the Coriolis force
acting on the gas (represented by the $\Omega^2$) stabilises long
wavelengths, so that instability only occurs on scales shorter than $\approx
2\pi G\sigma_{g0}/\Omega^2$ (as in classical GI). The
stabilising effect of \emph{gas}
pressure, on the other hand, does not make an appearance. 
In summary, instability occurs on a range of intermediate scales.
 But for
this range of unstable wavelengths to exist, the dust pressure must be
sufficiently weak or else the last term in \eqref{asymp2f1} swallows
the first term. 

Expression \eqref{asymp2f1} can be further simplified if we restrict
our attention to lengthscales much shorter than the gas's Jeans length
$\sim c_g^2/(G\sigma_{g0})$, but not so short that the dust pressure
dominates the other processes. In other words, we require
$ G\sigma_{g0}/c_g^2 \ll k \ll G\sigma_0/c^2$. If this intermediate
range exists, then on it 
 we may approximate
$\overline{\omega}_g^2\approx k^2c_g^2$ in the denominator of 
 the
first term in \eqref{asymp2f1}, and obtain
\begin{equation}
s \approx \frac{\Omega^2-\overline{\omega}^2}{\Omega}\,\epsilon - \delta\epsilon\Omega.
\end{equation}
Assuming further than $\delta\ll 1$ the last term may be dropped and
we have an expression similar to the single fluid one,
\eqref{epl}. The maximum growth rate
is easy to compute:
\begin{equation} \label{maxs1}
s_\text{max} \approx \frac{\epsilon\Omega}{Q^2},
\end{equation}
using the Toomre parameter for the dust. This expression is consistent
with Eq.~\eqref{1fwcs}, for
sufficiently large $Q$. But note that it only holds if there is a
sufficient separation of scales between the dust and gas pressure
scale heights, which is assured if $\eta\ll \delta$. 
 The wavelength of maximum growth is
also similar to that in a single fluid $\sim Q H$. 

\begin{figure}
   \centering
   \includegraphics[width=9cm]{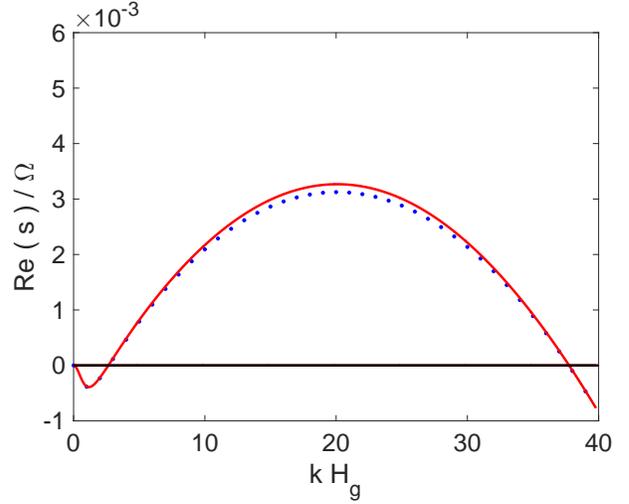}
   \caption{\label{fig3} Growth rate of the SGI as a function of
     radial wavenumber $k$ for weakly coupled particles. 
       Parameters are $Q_g=5$, $\epsilon=0.1$, $\delta=0.01$, and
     $\eta=10^{-4}$. Here $H_g= c_g/\Omega$ is the gas vertical scale
     height. 
      The solid red line is the full solution to the dispersion
     relation and the blue dotted solution is the asymptotic growth
     rate from \eqref{asymp2f1} }
\end{figure}

In Fig.~\ref{fig3} we plot the asymptotic growth rate as a
function of $k$ alongside the full solution obtained numerically from
\eqref{2fbig}. For the parameters chosen the agreement is respectable,
but naturally worsens once we leave the asymptotic regime of small
$\delta$, $\eta$, and $\epsilon$. Note that the lengthscale of fastest
growth is less than the gas-disk thickness $\sim  (\pi/10) H$, though
much longer than the particle-disk thickness because $H/H_g \sim
10^{-2}$.

\subsection{Pressureless turbulent dust}

Having treated the non-turbulent case, more relevant for weakly
coupled particles,
we now turn to a pressureless dust suspended in a turbulent gas disk,
the regime that best describes small dust. Thus in Eqs
\eqref{2flin1}-\eqref{2flin2} we set $D\neq
0$, but $c=0$, and it is assumed that $\epsilon \gg 1$. 
In addition, the turbulent diffusion is presumed small, $\alpha \ll 1$.
The simplest way to capture the full effects of diffusion is to let
$D k^2 \sim \epsilon^{-1}\Omega$, which (whatever the value of $D$)
will be true on some radial lengthscale. On longer wavelengths
persisting with this
scaling means we include harmless subdominant terms, while on shorter
wavelengths we expect diffusion to quench instability in any case. 
The resulting
equations
correspond exactly to those treated in Sections 2 and 3 in TI, 
which we now analyse in more detail and give
explicit expressions for the growth rates. 

\subsubsection{Instability criterion}

When $D\neq 0$, the onset of instability is difficult to calculate 
from the dispersion relation because unstable modes possess (small) complex
frequencies. Some progress can be made, however, in the limit of $\epsilon \gg
1$ and assuming that $s = \epsilon^{-1}s_1+\dots $ (thus extracting only SGI modes) and
$\alpha_g=\epsilon^{-1}\alpha_1+\dots$. Recall from Section 2.2 that in the tight coupling limit
$\alpha\approx \alpha_g$. 
To leading order the dispersion relation \eqref{2fbig} becomes the
quadratic
\begin{align}
&(1+\delta)\left[(1+\delta)\mu^2-\delta c_g^2 k^2\right]\Omega^2s_1^2
  -\left\{\delta c_g^2(\mu^2-c_g^2k^2) \right. \notag \\
& \hskip1cm
\left. +\alpha_1(1+\delta)\left[(1+\delta)\overline{\omega}^2-\delta
    c_g^2k^2\right]  \right\}k^2c_g^2\Omega s_1 \notag \\
&\hskip2cm  + \delta c_g^4k^4\alpha_1\Omega^2 =0. \label{quad}
\end{align}
In Eq.~(13) TI present an equivalent expression.
For reasonable values of $\delta$ and $Q_g>1$
the coefficient of $s_1^2$ is positive and so the sign of the growth
rates can be determined from the coefficient of $s_1$. After some
manipulation the instability criterion is
\begin{equation}
Q_g < [\delta + \epsilon\alpha_g(1+\delta)]
\sqrt{\frac{1+\delta}{\epsilon\alpha_g
\left[\delta+\epsilon\alpha_g(1+\delta)^2\right]}},
\end{equation}
which agrees with Eq.~(18) in TI.
Taking small $\delta$, this simplifies to
\begin{equation} \label{stab2}
Q_g \lesssim \sqrt{1+ \frac{\delta}{\epsilon\alpha_g}},
\end{equation}
which is remarkably similar to the diffusionless criterion 
\eqref{stab1}. The smaller the turbulent diffusion $\alpha_g$,
the greater the range of instability. Diffusion on small scales
has replaced dust pressure in \eqref{stab1} in stabilising the 
secular GI, but the two processes work exactly the same otherwise. 
Finally, note that \eqref{stab2} differs from the final criteria in 
TI because they make additional and unnecessary assumptions regarding
the relative sizes of $\epsilon$, $\delta$, and $\alpha_g$.

\subsubsection{Asymptotic growth rates}

Solutions to the quadratic \eqref{quad} are ugly, but if
we take the limit $\delta\ll 1$,  we obtain
\begin{align} \label{asymp2f2}
\text{Re}(s) = \frac{1}{2}\left(\frac{\sigma_0}{\sigma_{g0}}\right)
\left(\frac{2\pi G\sigma_{g0}|k| -
 \Omega^2}{\epsilon\,\overline{\omega}_g^2}\right)
\left(\frac{k^2c_g^2}{\Omega}\right) - \frac{1}{2}k^2\alpha.
\end{align}
This is almost identical to expression \eqref{asymp2f1}. 
The main difference is that the
stabilising term on short scales arises
from turbulent diffusion, rather than dust
pressure, once again. The long wavelength stabilisation is the same --- the Coriolis
force. Instability can only occur if these two stabilising scales are
sufficiently well separated. 

\begin{figure}
   \centering
   \includegraphics[width=9cm]{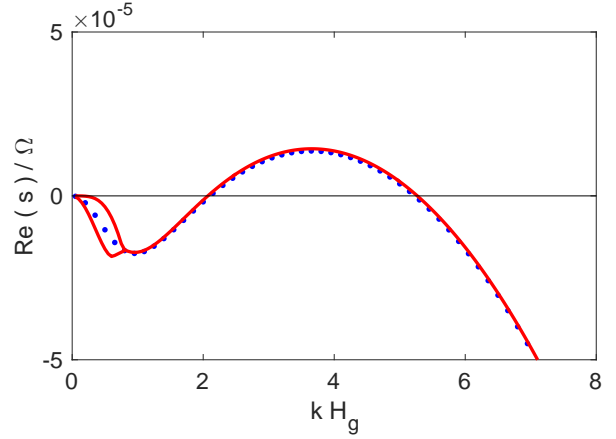}
   \caption{\label{fig4} Growth rate of the SGI as a function of $k$
     for tightly coupled particles. Parameters are: $Q_g=3$, $\epsilon=100$, $\delta=0.01$, and
     $\alpha=10^{-5}$.
The solid red line is the full solution to the dispersion relation, and
the blue dots correspond to the asymptotic growth rate \eqref{asymp2f2}.}
\end{figure}

In Fig.~\ref{fig4} we plot the asymptotic growth rate
\eqref{asymp2f2} as a function of $k$ next to the full solution
obtained numerically from \eqref{2fbig}. For a representative set of
parameters the agreement is good, though weakens the smaller
$\epsilon$ and the larger $\delta$, as expected. Note also that at
small $k$ the approximation fails to capture the bifurcation
 into two monotonically growing modes with different growth
rates. Finally, the wavelength of fastest growth is $\sim H_g$, and thus the
approximation of a razor thin disk is only marginally applicable.

The approximate maximum growth rate can be obtained by a similar
argument to that given in Section 5.3.2. 
We find to leading order on the intermediate 
range $ G\sigma_{g0}/c_g^2 \ll k \ll
G\sigma_0/(\epsilon \alpha\Omega)$ and for $\delta\ll 1$ that 
\begin{align}\label{pressurelesslimit}
s\approx \frac{1}{2}\frac{\pi G \sigma_0|k|}{\epsilon\Omega}- k^2\alpha.
\end{align}
This expression is similar to the growth rate in the single-fluid
pressureless case, \eqref{pless1f}, differing only by a half (on
account of the unstable mode appearing as a complex conjugate pair). 
The maximum growth rate is staightforward
to compute, and we find
\begin{align}\label{maxys}
\text{Re}(s_\text{max}) = 
\frac{1}{2}\frac{\pi^2 G^2\sigma_d^2}{\epsilon^2\alpha} 
 =\frac{1}{2}\frac{\Omega}{\epsilon^2\alpha_g Q_g^2},
\end{align}
with maximum growth occuring on lengthscales $\sim \epsilon \alpha_g
Q_g H_g$(as in the single fluid).
Note that a similar expression to \eqref{pressurelesslimit} (and
identical to \eqref{pless1fmax}) can also be
achieved by enforcing the terminal velocity approximation for the dust
in Eq.~\eqref{2flin1b}, and then radial geostrophic balance for the
gas in Eq.~\eqref{2flin1c}. The latter assumption, combined with the
remaining equations, ensures that
$U_x'\sim \delta u_x'$ and so, to leading order, the growth rate is
the same as in the single fluid treatment.

\subsection{Physical picture}

To complete this section we bring together the insights gained from
the preceding analyses and construct a physical model for how the
secular GI works in a two-fluid disk. As is clear from the asymptotic
estimates \eqref{asymp2f1} and \eqref{asymp2f2}, instability occurs
only when there exists a range of
intermediate wavelengths in which both the gas's tidal forces and 
dust pressure/turbulence are sufficiently weak. 
Obviously, dust pressure or turbulence
will interfere with gravitational collapse on small scales, stopping
the clumping of material into rings. Instability
hence migrates to longer scales where this effect is weak. On sufficiently
long scales, however,
the tidal force will send a disturbed gas parcel into 
epicyclic motion, which also impedes clumping. 
To counteract this
effect, 
instability
must then move to shorter scales where the gas pressure is sufficiently
strong to block the gas's epicyclic tendency (via a zonal,
 or geostrophic, flow). 
Criteria \eqref{stab1} and \eqref{stab2} describe necessary
conditions that permit some band of wavelengths to satisfy these two
requirements.  

Essentially, the two-fluid mode is attempting to be as close to the
single-fluid model (Section 3) 
or the two-fluid incompressible model (Section 4) as possible. 
On this intermediate range of
lengthscales two dust blobs radially displaced towards one another
exchange their angular momentum with the background gas, and hence can
gravitationally collapse rather than be sent into epicycles. The
gas, however, being no longer an infinite reservoir of angular
momentum undergoes a commensurate perturbation. If the gas
pressure is sufficiently strong, however, this perturbation takes the form of a
zonal flow, not an epicycle, and so the structure of the instability is
retained. If we move to longer and longer scales, the gas pressure
weakens and so can only support very mild zonal flows, unable to carry
appreciable angular momentum from the drifting dust. As a consequence,
the instability is suppressed. On the other hand, whereas the gas is in radial geostrophic balance,
the dust can fall into either geostrophic balance (for
$\epsilon\ll 1$) or terminal velocity balance (for $\epsilon \gg 1$). 

\section{Discussion}

In this section we apply the stability criteria and growth rate
estimates of the two-fluid model to reasonable models of
protoplanetary disks. Because of uncertainties in the parameters and
deficiencies in the model itself, we do
not attempt to be comprehensive or to hold fast to the quantitative
results we obtain. Rather the aim is to give a sense of the main
trends and qualitative behaviour, and the general range of numbers one
might find in a real system. To that purpose we first concentrate on
the two populations of strongly and marginally coupled particles,
using the simple estimates derived previously. We then solve the full
dispersion relationship numerically at each radius. 

The disk model we use is a variant of the minimum mass solar nebula
(see Youdin 2011).
The background gas disk surface density is given as a function
of disk radius by
\begin{equation} \label{MMSN}
\sigma_0 = 2\times 10^3 F R_\text{AU}^{-3/2} \,\text{g cm}^{-2}.
\end{equation}
Here $F$ is a
free dimensionless parameter, and $R_\text{AU}$ is disk radius in
units of AU. The temperature of the nebula is given by 
\begin{equation}
T = 200\,R_\text{AU}^{-1/2}\,\text{K}.
\end{equation}
As a consequence, the gas's Toomre parameter is
\begin{equation}
Q_g \approx \frac{40}{F} R_\text{AU}^{-1/4}.
\end{equation}
Thus at 1 AU, $Q_g$ is roughly 40 and this falls to about 10
as we approach 100 AU. 
The inverse Stoke's number can be approximated by
\begin{equation}
\epsilon = 4\times 10^3 \frac{F}{a_\text{mm}}R_\text{AU}^{-3/2},
\end{equation}
where $a_\text{mm}$ is particle size in units of mm. 
We have assumed that the particles always lie in the Epstein
regime. Only the largest particles at the smallest radii enter the
Stokes drag regime, so to make life simple we omit it. As a result, stability
can be determined once $\delta$, $F$, $\alpha_g$, particle size,
and the disk radius 
are specified.

\subsection{Weakly coupled particles}

First consider larger particles with a largish Stokes number
$\epsilon \lesssim 1$. From Figure 1, in a standard MMSN 
these might correspond to $\sim$ 10 cm
particles at 30 AU or more or $\sim 1$ cm particles at 100 AU. 
In an older less massive disk, this class may also include mm
particles but only at 100 AU. Thus most particles do not fall into
this regime. 

Secular GI
arises
when criterion \eqref{stabbers1} is fulfilled. Taking the standard
value for the dust to gas ratio $\delta = 10^{-2}$, we only need to
estimate the ratio of velocity dispersions. Assuming that the
particles'
random motions are controlled by the background turbulence, as in
Section 2, we have $c/c_g \sim \sqrt{\epsilon \alpha_g}$ and thus
instability occurs when
\begin{equation}
Q_g < Q_\text{crit} \approx 10^{-1} \epsilon^{-1/2}\alpha_g^{-1/2}.
\end{equation}
This criterion includes both the classical GI of a dust layer and the
secular GI, in which we are more interested. 
As discussed earlier, the properties of the turbulence
 are poorly constrained. We thus allow
$\alpha_g$ to vary between $10^{-7}$ (a perhaps unrealistically low
level) 
and
$10^{-3}$. 
Next, to fix ideas, we set $\epsilon\sim 0.1$  and find that the critical
Toomre parameter for the gas is
$$ Q_\text{crit} \approx  10-10^3,$$
with the lower value corresponding to a thick disk of relatively `hot'
particles ($\alpha_g=10^{-3}$),
and the higher value to a thin disk of colder particles
($\alpha_g=10^{-7}$).

Our standard MMSN with $F=1$ yields $Q_g$ that fall directly in this
range. If $\alpha_g=10^{-3}$ instability is not possible, but if
$\alpha_g\geq 10^{-4}$ then instability can occur on most radii.
 Less
massive disks exhibit larger $Q_g$; for instance with $F=0.1$, we have
$Q_g>100$ at all radii, and thus the existence of SGI is very much
 conditional on the efficiency of the turbulence. If $\alpha_g\sim
 10^{-7}$ then cm sized particles, or even smaller, could be unstable in
 the outer regions of low mass disks. 

What of the growth rates of the unstable modes? Equation \eqref{maxs1}
gives an upper bound on $s$, in the regime of larger particles.  This
can be reworked into
\begin{equation}
s_\text{max} \approx \delta^{2}Q_g^{-2} \alpha_g^{-1}\Omega \sim 10^{-6}\alpha_g^{-1}\Omega,
\end{equation}
where the last equality comes by setting $\delta=10^{-2}$ and
$Q_g\sim 10$. For relatively strong turbulence $\alpha_g \sim
10^{-4}$, the e-folding time of an unstable mode is $10^2$ orbits, 
too long to be relevant at 100 AU, but possibly significant at smaller
radii, for example 10 cm size particles at $\sim 10$ AU. 
Smaller $\alpha_g$, of course, yield faster growth 
on relevant timescales.

\subsection{Tightly coupled particles}

We next consider well coupled particles, a class that covers most
solids of interest (see Figure 1).
 The relevant stability criterion for this
dust is given by \eqref{stab2}. The dimensionless 
diffusion coefficient for tightly
coupled particles is $\alpha\approx \alpha_g$. 
Setting $\delta=0.01$ and $\epsilon=10$
yields the criterion
\begin{equation}
Q_g \lesssim \sqrt{1+10^{-3}\alpha_g^{-1}},
\end{equation}
which is more difficult to satisfy than in the weakly coupled case. 
If we assume
that $\alpha_g = 10^{-7}-10^{-3}$, the condition becomes
$$ Q_g < Q_\text{crit} \approx 1-100,$$
where the larger value corresponds to inefficient turbulent
diffusion ($\alpha_g=10^{-7}$) and the lower bound corresponds to
efficient diffusion ($\alpha_g=10^{-3}$).
Once again, this suggests that instability occurs when the turbulence is
sufficiently weak. In fact, given $F=1$, the SGI is suppressed if
$\alpha_g \geq 10^{-5}$. The situation worsens when $\epsilon>10$.
The conclusion is that the instability may not 
be widespread in smaller dust.

Let us next turn to growth rates, in particular expression
\eqref{maxys}. To fix some numbers, we generously set
$\epsilon=10$ and $\delta=0.01$ and after some manipulation obtain
$$ \text{Re}(s_\text{max}) \sim 10^{-6} Q_g^{-2}
\alpha_g^{-1}\Omega.$$
If $Q_g\sim 10$, then
we have 
\begin{equation}
 \text{Re}(s_\text{max}) \sim 10^{-8}\alpha_g^{-1}\Omega.
\end{equation}
While $\alpha_g \leq 10^{-7}$ yields appreciable growth at all radii,
$\alpha_g= 10^{-6}$ does so only for $R\lesssim 10$ AU.
For larger $\epsilon$ growth times lengthen.
This further reinforces the conclusion that the SGI is only
relevant to the dynamics of tightly coupled
particles when turbulence is very low indeed.

\subsection{Marginally coupled particles 
                     in realistic disk models}

\begin{figure*}
   \centering
  \includegraphics[width=15cm]{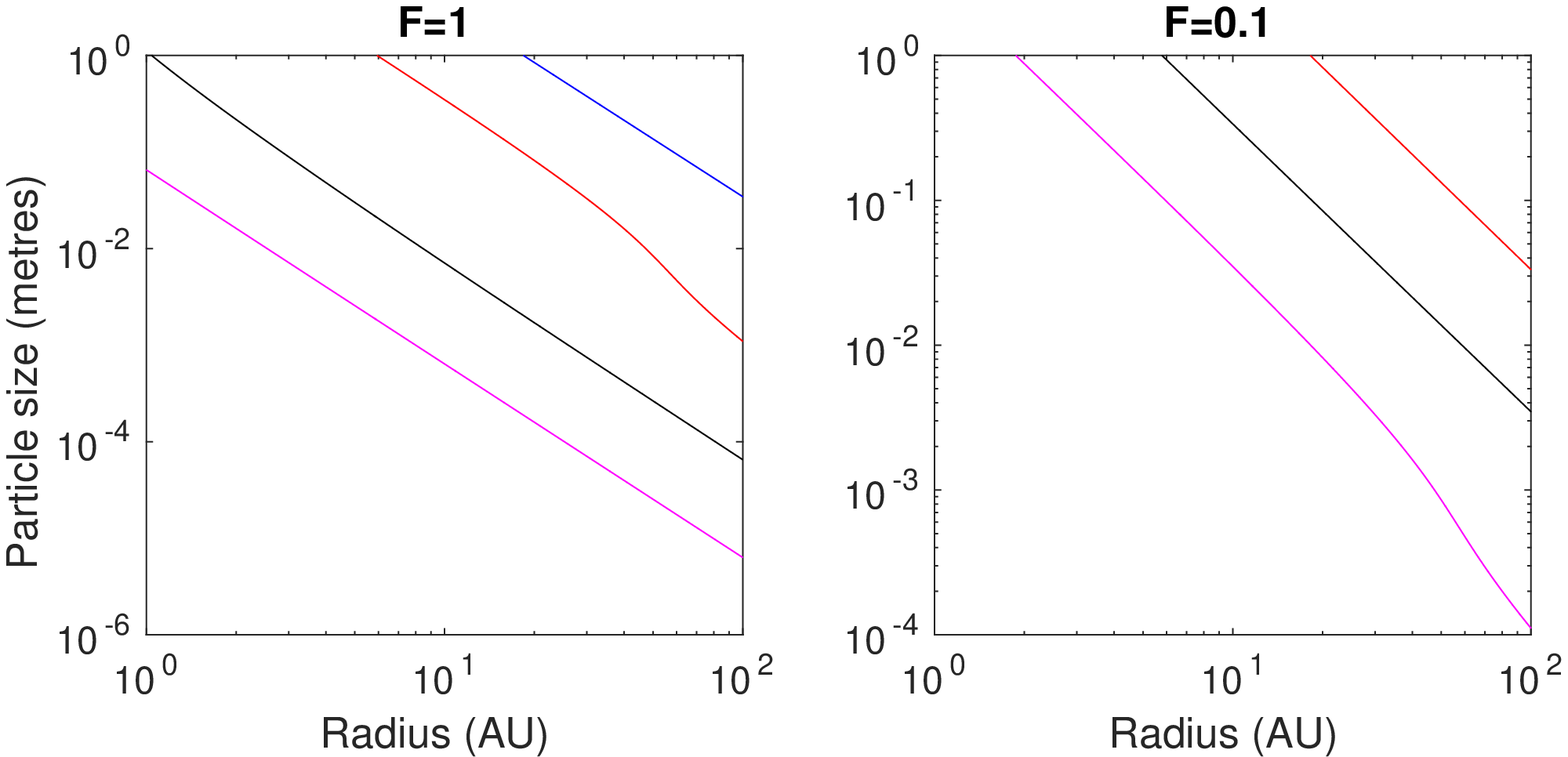}
  \includegraphics[width=15cm]{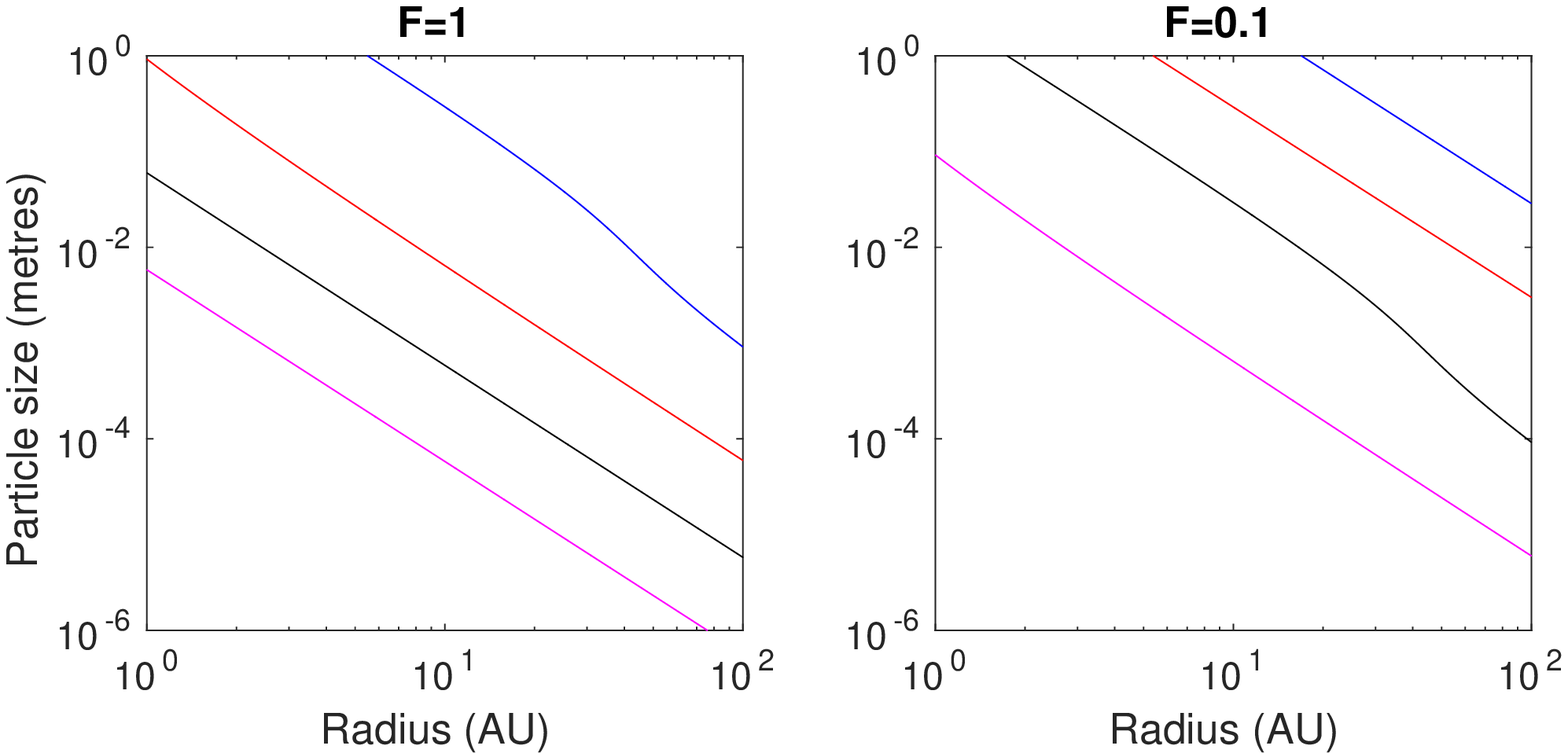}
   \caption{\label{fig5} Stability curves in a plane of disk radius
     and particle size. The pink, black, red, and blue curves
     correspond to $\alpha_g=10^{-7},\,10^{-6},\,10^{-5},\,10^{-4}$. 
     The regions above the respective curves are
     unstable. Two disk models are employed, a more massive disk
     with $F=1$ (left panels) and a
     less massive disk with $F=0.1$ (right panels). The upper row
     corresponds to a dust-to-gas
     mass ratio of $\delta=0.01$, while the lower row to $\delta=0.1$. }
\end{figure*}

In this subsection we solve the full dispersion relation at each
radius of our disk model. 
Figure 1 indicates that particle sizes of a mm and above couple to
the gas differently at different radii, 
potentially passing from the well-coupled to the weakly coupled
regimes as we go further out radially. At certain radii
$\epsilon\sim 1$ and our analytic results are no longer strictly
valid, meaning that the dispersion relation \eqref{2fbig} 
must be solved numerically.

Some stability curves are plotted in Fig.~\ref{fig5} for two different
disk models, $F=1$ (left panels) and $F=0.1$ (right panels), 
for several values of the turbulence parameter 
$\alpha_g$, and for two values of the dust to gas density ratio,
$\delta=0.01$ (top row) and $0.1$ (bottom row).  
Parameter regions above a given curve are subject
 to instability, and thus for a given $\alpha_g$ and particle size 
there exists a critical radius within which the SGI is completely suppressed.
In fact, for a
relatively turbulent disk with $\alpha_g=10^{-4}$, $F=1$ and $\delta=0.01$,
 all particles smaller than $\approx 3$ cm are stable,
 and all particles smaller than 1 mm, when $\alpha_g=10^{-5}$. Weaker
 levels of turbulence, of course,
 permit instability upon smaller particle sizes and for larger swathes
 of the disk. But one must drive $\alpha_g$ to levels $\sim 10^{-7}$
 to obtain SGI at radii $< 10$ AU for particles larger than a mm. 

The SGI's struggles worsen
 when the disk is less massive ($F=0.1$),
 with mm sized particles
 unstable only for very low values of $\alpha_g$. But moving to the lower
 row of plots, it is immediately clear that increasing $\delta$
 improves its range. For example, if $F=1$ and
 $\alpha_g=10^{-4}$ then all particles smaller than mm 
sizes are stable.
If $\alpha_g=10^{-5}$ then all particles smaller than $\sim 0.1$ mm
are stable. On the
other hand, when $\delta<0.01$ the prospects for instability become
increasingly bleak. 

\begin{figure}
   \centering
  \includegraphics[width=9cm]{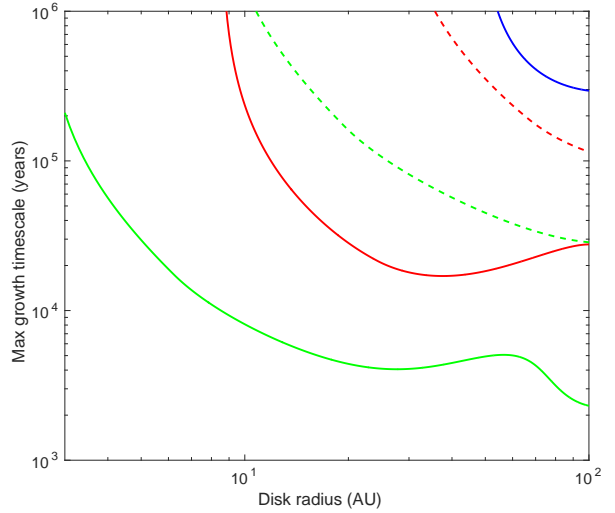}
   \caption{\label{fig6} Maximum e-folding times for the SGI at
     different radii in a standard disk model with $F=1$ and
     dust-to-gas ratio $\delta=0.01$. The solid
     curves correspond to cm sized particles, and the dashed curves to
   mm sized particles. The blue colour represents a turbulent mass
   diffusivity of $\alpha_g=10^{-5}$, red represents
   $\alpha_g=10^{-6}$, and green $\alpha_g=10^{-7}$.}
\end{figure}
We also compute the minimum e-folding times
for the SGI, for $F=1$ and
$\delta=0.01$. These are plotted in Figure \ref{fig6}.
Solid curves correspond to cm sized
particles and dashed curves correspond to mm sized particles. 
We omit smaller particles because they are always in the
well-coupled regime treated in Section 6.2. The different colours
represent different values of $\alpha_g$. As is clear, a turbulence
level of $\alpha_g> 10^{-5}$ yields a growth time too long to be
important for both particle classes, while a value of
$\alpha_g=10^{-6}$ yields a growth time of a few $10^4$ years for cm
sized particles, at a large range of radii $\gtrsim 20$ AU. Milimetre
sized particles exhibit appreciable growth only for low turbulence
levels $\alpha_g\sim 10^{-7}$ and at larger radii $\gtrsim 50$
AU. Less massive disks (such as with $F=0.1$) yield even weaker
growth. Increasing $\delta$ to 0.1,
exacerbates growth by up to an order of magnitude, whereas decreasing
$F$ to 0.1, reduces the growth rate by roughly an order of magnitude.

These growth times are on the whole consistent with Youdin
(2011) at radii $\gtrsim 20$ AU, but closer to the disk two-fluid
effects lead to noticeable departures. The growth times diverge at
certain critical radii. These occur, of
course, when the SGI is stabilised, as indicated by Fig.~\ref{fig5}.
For a given $\alpha_g$, any given particle size has a critical radius
within which the SGI is stable. The takeaway message is that mm sized
particles require very low levels of turbulence to be SGI unstable,
and then this is localised to the outer parts of the disk. Centimetre
sized particles do better, and the SGI may play some role in their
dynamics across a range of disk radii and disk properties. Finally,
older, less massive disks struggle to host the SGI in any form, though
a larger dust to gas ratio can mitigate this to some
degree. Unfortunately, like $\alpha_g$, the parameter $\delta$ is
difficult to constrain.

\section{Conclusion}

In this paper we have explored the secular gravitational instability (SGI)
using a simple two-fluid model. Despite the complexity of its
associated sixth order dispersion relation, analytic stability
criteria and growth rates can be obtained in the two limits of weakly
and strongly coupled particles. We find that on sufficiently long and short
radial scales the SGI is stabilised; the existence of an unstable
range of intermediate scales leads to an explicit instability
condition involving the gas's Toomre parameter, 
a distinctive feature of the two-fluid SGI, as opposed to the
single-fluid version.

The mathematical analysis suggests a straightforward way to understand
the instability mechanism. The SGI favours intermediate scales upon
which stabilising dust pressure or turbulence is weak, but upon which the gas
pressure is strong. The latter condition permits the gas to fall into
geostrophic balance: hence when the gas is azimuthally accelerated by the
dust drag, it will form a zonal flow rather than undergo epicycles
that would disrupt the radially collapsing dust. 

An assessment of the prevalence of SGI in real disk models is handicapped
by uncertainties in two parameters, the strength of the turbulence
$\alpha_g$, and the mass ratio of a certain species of dust to the gas
within the dust subdisk, $\delta$. Starting with a fiducial value of
$\delta=0.01$, we find that a moderate level of turbulence
$\alpha_g=10^{-5}$ prohibits the SGI on most radii, and when it does
occur it grows too slowly $\sim 10^5$ years --- the timescale of the
large-scale evolution of the disk, and of appreciable radial drift. Weaker
turbulence $\alpha_g=10^{-6}$ permits growth for cm sized particles on
radii $\gtrsim 10$ AU, with efolding times of a few $10^4$
years. Smaller sized particles may be subject to SGI but grow too
slowly. It is only for $\alpha_g=10^{-7}$ that mm sized particles
sustain growth at reasonable levels, and then only for $R> 10$ AU. 
Increasing $\delta$ improves the situation, of course, and
$\delta>0.01$
might
be the case for particularly well-settled and populous subclasses of
particle, though further work is needed to better constrain this
parameter. Even so, if $\alpha_g>10^{-6}$ it may be prove difficult for
the SGI to meaningfully impose itself on the disk dynamics.

We also discuss the various shortcomings of the
razor-thin model we employ, which is especially a problem when the
dust and gas disks exhibit different scale thicknesses. 
These issues no doubt impact quantitatively on
our results, but the main qualitative conclusions and our picture of
instability should be
robust. They can be checked with a suitable vertically stratified
analysis akin to Mamatsashvili \& Rice (2010) and Lin (2014), which
will also provide more reliable quantitative estimates on the stability
curves and growth rates.

Our results extend previous analyses of the SGI, and for larger radii
are in relative agreement with Youdin (2011) and Shariff \& Cuzzi
(2011). A notable difference is that the two-fluid model prohibits SGI
on radii less than a critical radius. As a result, the SGI is
certainly unviable on radii $< 1 $ AU, and possibly absent on radii
$<10$ AU, the expected regions of planet formation. The prospects for
SGI in the cm class of particles on disk radii $\sim 10$ AU 
are reasonable as long as gas turbulence is not too efficient
$\alpha_g \lesssim 10^{-6}$. The instability could then be an
important route by which large aggregates could form further out, 
leapfrogging the
entire range of difficult cm to km sizes. Note that our results are
only for axisymmetric instability. It is likely, via analogy with
classical GI, that non-axisymmetric SGI occurs for larger $Q_g$, in
which case our stability curves may need some revision.

It has been
hypothesised that the SGI generates observed 
dust ring structures at larger
radii in protoplanetary disks (TI). 
As discussed in Section 6, however, the SGI has great difficulty on
radii $\gtrsim 10$ AU for small particle less than a cm in size. The
dust-to-gas ratio $\delta$ needs to be increased, and $Q_g$
taken to levels approaching 1 in order to obtain instability. While it may be
possible to justify increasing $\delta$, such a low $Q_g$ would mean
the gas disk is marginally unstable to classical GI.
Perhaps a more important point is that, while the linear phase of
the SGI evolution is axisymmetric, its nonlinear phase will most likely
involve a non-axisymmetric breadown into disordered flow,
as in classical GI, not the formation of large-scale quasi-steady rings.
Dedicated nonlinear simulations are required to test what dynamics
the SGI exhibits once it reaches nonlinear amplitudes, and how readily
it forms planetesimal clumps. This forms the basis of future work. 

\section*{Acknowledgments}

The authors thank the reviewer, Dick Durisen, for a set of 
comments that led to a much improved manuscript. HNL acknowledges
partial funding from STFC grant ST/L000636/1, and 
RR from a Bridgewater summer internship and from Newnham college.

\end{document}